\documentclass{iaus}
\usepackage{graphicx}

\newcommand{\msun}{\mbox{$M_{\odot}$}}

\newcommand{\zsun}{\mbox{$Z_{\odot}$}}
\newcommand{\teff}{\mbox{$T_{\rm eff}$}}
\newcommand{\Teff}{\mbox{$T_{\rm eff}$}}

\newcommand{\vinf}{\mbox{$v_{\infty}$}}
\newcommand{\vesc}{\mbox{$v_{\rm esc}$}}

\newcommand{\mdot}{\mbox{$\dot{M}$}}

\newcommand{\msunyr}{\mbox{$M_{\odot} {\rm yr}^{-1}$}}

\newcommand{\kms}{km s$^{-1}$}

\title[Stellar winds]
{Mass loss and evolution of hot massive stars}

\author[Vink,J.S.]  
{Jorick S. Vink$^1$}

\affiliation{$^1$Armagh Observatory, College Hill, Armagh, BT61 9DG, Northern Ireland, United Kingdom 
\break email: jsv@arm.ac.uk}

\pubyear{2008}
\volume{252} 
\pagerange{}
\date{?? and in revised form ??}
\jname{Proceedings Title IAU Symposium}
\editors{L. Deng, K.L. Chan \& C. Chiosi, eds.}
\begin{document}

\maketitle

\begin{abstract}
We discuss the role of mass loss for the evolution of the most massive stars, highlighting the 
role of the predicted {\it bi-stability} jump that might be relevant for the evolution of rotational 
velocities during or just after the main sequence. This mechanism is also proposed as an 
explanation for the mass-loss variations seen in the winds from Luminous Blue Variables (LBVs). These 
might be relevant for the quasi-sinusoidal modulations seen in a number of recent transitional 
supernovae (SNe), as well as for the double-throughed absorption profile recently discovered in the 
H$\alpha$ line of SN 2005gj. Finally, we discuss the role of metallicity via the $Z$-dependent
character of their winds, during both the initial and final (Wolf-Rayet) phases of evolution, with 
implications for the angular momentum evolution of the progenitor stars of long gamma-ray bursts (GRBs).

\end{abstract}

\firstsection 

\section{Introduction}

Mass loss has a major effect on the evolution of stars of all initial masses, however its effect is most 
prominent for the more massive stars due to their large luminosities -- in close proximity to the 
Eddington limit. Mass loss is relevant both in terms of evolutionary pathways as well as the properties 
of the pre-supernova (SN) circumstellar environments. In this contribution, we first discuss mass-loss 
predictions that are relevant for predicting the forward evolution of massive stars. 
There are actually two aspects that need to be accounted 
for: (i) the loss of {\it mass} as winds ``peel off'' the star's outer layers (Conti 1976), but 
as massive stars start their evolution as rapid rotators, also 
(ii) the associated loss of {\it angular momentum} (e.g. Langer 1998). 

Towards the end of the main sequence massive stars encounter the so-called bi-stability jump, for which we 
discuss the implications of the loss of angular momentum.
For the final stages, the evolution of angular momentum is particularly relevant for our understanding of the 
long gamma-ray burst (GRB) phenomenon, as the popular collapsar model (MacFadyen \& Woosley 1999) 
requires the core of the progenitor star to be rapidly rotating before collapse (but see also Lee \& Ramirez-Ruiz 2006). 
A key parameter in the story is that of metallicity, due to   
the $Z$-dependence of radiation-driven winds during both the main sequence and evolved
Luminous Blue Variable (LBV) and Wolf-Rayet (WR) phases of massive star evolution.

In the canonical scenario of massive stars, LBVs represent a transitional phase between 
the main sequence and the WR phase, however we also discuss the possibility for a new evolutionary paradigm
in which the variable winds of LBVs might betray themselves as the {\it direct} progenitors of SNe.

\section{Mass loss predictions}
\label{sec:massloss}

The evolution of a massive star, with $M$ $>$ 30 \msun\ is largely determined by 
the strength of their winds, which depends on the luminosity ($L$), mass ($M$), and
metallicity $Z$. 
The Fe-group elements are particularly efficient scatterers of photons at specific 
line frequencies responsible for the amount of mass loss, whilst the CNO elements set
the wind terminal velocity (Vink et al. 1999). 
There are currently two basic methods in use for predicting the mass-loss rates from 
massive stars, which have been reviewed in detail by Vink (2006). 
In short, the first method concerns the modified-CAK (Castor et al. 1975) method 
(Kudritzki \& Puls 2000), the second one involves the Monte Carlo approach 
(Abbott \& Lucy 1985, Vink et al. 2000). Both methods have their pros and cons. 
In the first approach, the wind hydrodynamics are 
more or less self-consistently solved for (albeit using depth-dependent force multiplier 
parameters), however multi-line scattering is not accounted for. This aspect is 
included in the second approach, where the line acceleration is calculated for all radii, 
although most Monte Carlo predictions do not 
properly account for the wind hydrodynamics (but see Vink et al. 1999). 

\begin{figure}
 \includegraphics[height=.5\textheight]{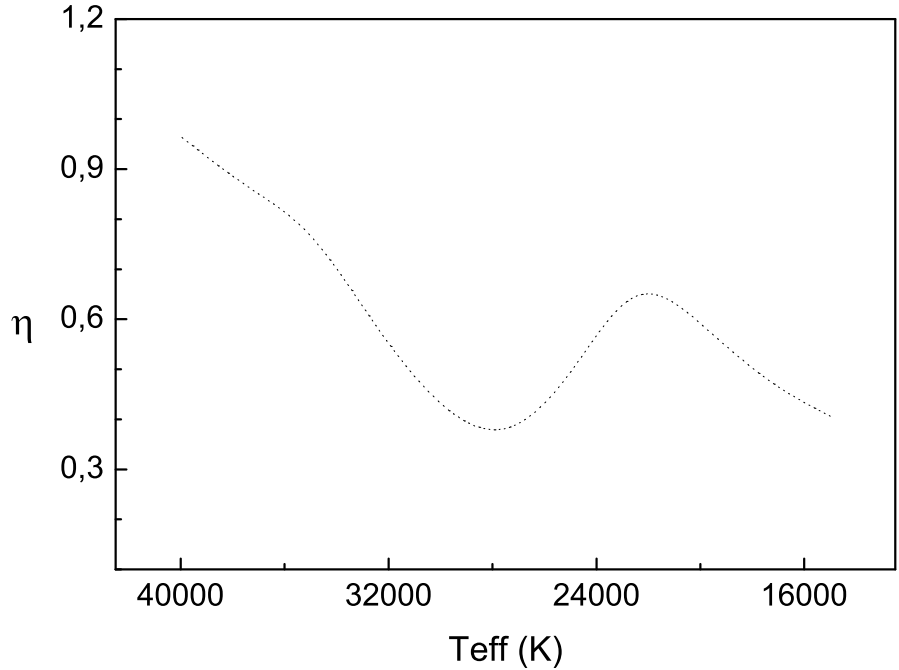}
  \caption{Predicted mass-loss behaviour as a function of effective temperature. Note the
position of the local maximum at about 22000 K, coincident with the position of the
empirical bi-stability jump at spectral type B1 (Lamers et al. 1995).}
\label{fig:vink99}
\end{figure}

Vink et al. (1999,2000) predicted the mass-loss rates of OB supergiants as a function
of the stellar parameters ($L$,$M$, and \Teff) including multiple scatterings on line and continuum
opacity with a Monte Carlo approach and their mass-loss rates were computed as a function 
of the wind {\it terminal velocity} -- a parameter accurately (within 10 \%) 
retrievable from ultraviolet P\,Cygni line profiles. 
This is in stark contrast to empirical {\it mass-loss rates} which are highly 
uncertain (by factors up to 3-10) due to uncertainties in wind ionization and wind 
inhomogeneities (see Hamann et al. 2008 for a
recent overview on the issue of wind clumping).
The Vink et al. mass-loss rates are found to scale as:

\begin{equation}
\dot{M}~\propto~L^{2.2}~M^{-1.3}~\Teff^1~(\vinf/\vesc)^{-1.3}
\label{eq_formula}
\end{equation}
showing that $\dot{M}$ depends rather steeply on the stellar luminosity ($L^{2.2}$).
The reason for this is that brighter stars have denser winds and Monte Carlo predictions  
yield an increasingly larger mass-loss rate than modified-CAK predictions.
The Vink et al. mass-loss recipes 
are widely used in models for massive star evolution 
(e.g. Meynet \& Maeder 2003, Limongi \& Chieffi 2006, 
Eldridge \& Vink 2006, Brott et al. in prep.).

Figure~\ref{fig:vink99} depicts how the predicted mass-loss rates vary as a function of 
$\teff$ and thus how a massive star may find its wind change during its coarse of 
evolution. The predictions are expressed in terms of the wind momentum efficiency, or wind 
performance number, $\frac{\dot{M} v_{\infty}}{L/c}$.
The figure shows a declining wind efficiency with $\teff$. 
At high temperatures ($\sim$ 40~000 K) the wind momentum is large due to the fact that 
the radiative flux and the opacity have a good ``match'' with respect to their wavelength distribution. 
However, when $\teff$ drops, the stellar flux moves away from its maximum towards lower (optical)
wavelengths, which results in an ever-growing mismatch between the flux and the ultraviolet line opacity.
At $\sim$ 25~000 K, a sudden mass-loss discontinuity is noted. This is due to an increased Fe 
opacity when Fe {\sc iv} recombines, and the more abundant Fe {\sc iii} lines provide most of the line force in the inner wind 
(Vink et al. 1999). This ``bi-stability jump'' (Pauldrach \& Puls 1990) may recently have been confirmed in 
radio data that appear to confirm the presence of a local maximum (Benaglia et al. 2007, Markova \& Puls 2008), however it should also be 
noted that the predicted values below the temperature of the jump appear to be much larger than those 
found from empirical modelling by up to a factor of 10 
(Vink et al. 2000, Trundle \& Lennon 2005, Crowther et al. 2006). 

The bi-stability jump where winds change from a low $\dot{M}$, fast wind, to a high $\dot{M}$, slow wind 
may comprise an important ingredient for stellar evolution calculations when stars evolve off 
their main-sequence positions towards the lower $\teff$ part of the Hertzsprung-Russell diagram (HRD). 
This is not only relevant for their {\it mass} loss, but also for the associated loss 
of {\it angular momentum}. 

\begin{figure}
 \includegraphics[height=.45\textheight]{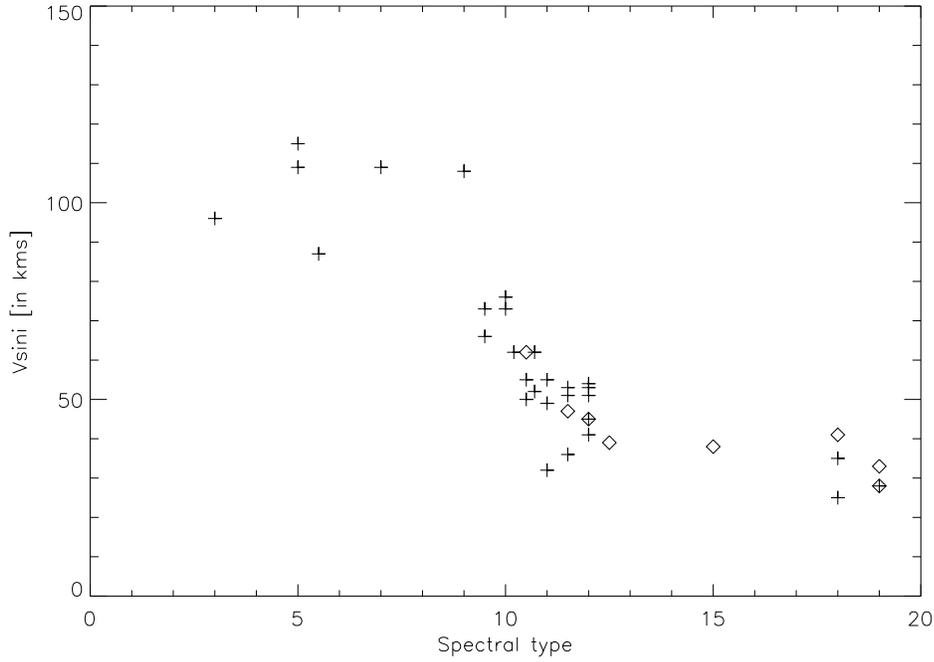}
  \caption{The projected rotational velocity, $v$sin$i$ of Galactic B supergiants 
as a function of spectral type, starting with O3 to O9 and then switching to B0 (at 10) and to A0 (at 20). 
It can be noted that $v$sin$i$ drops from $\sim$100 \kms\ to $\sim$50 \kms\ at around spectral type B0, 
which is close to the spectral type of the bi-stability jump B1 (Lamers et al. 1995). The figure has been adapted
from Markova \& Puls (2008).}
\label{fig:mp}
\end{figure}

\section{Angular momentum loss}

Massive stars rotate rapidly at birth (with $v_{\rm rot}$ $\simeq$200-300 \kms) and remain relatively 
rapid rotators throughout their main-sequence lifetimes. Obviously, $v_{\rm rot}$ decreases due to 
the angular momentum loss via stellar winds, which
implies that the effects are largest at the highest initial masses and luminosities, and metallicities.
Furthermore, when the objects evolve off the main sequence, they swell up to become (super)giants, and 
$v_{\rm rot}$ is anticipated to drop due to the increase in stellar radius (e.g. Hunter et al. 2008). However, is this the 
entire story or is the bi-stability jump also of relevance?

\begin{figure}
 \includegraphics[height=.65\textheight,angle=270]{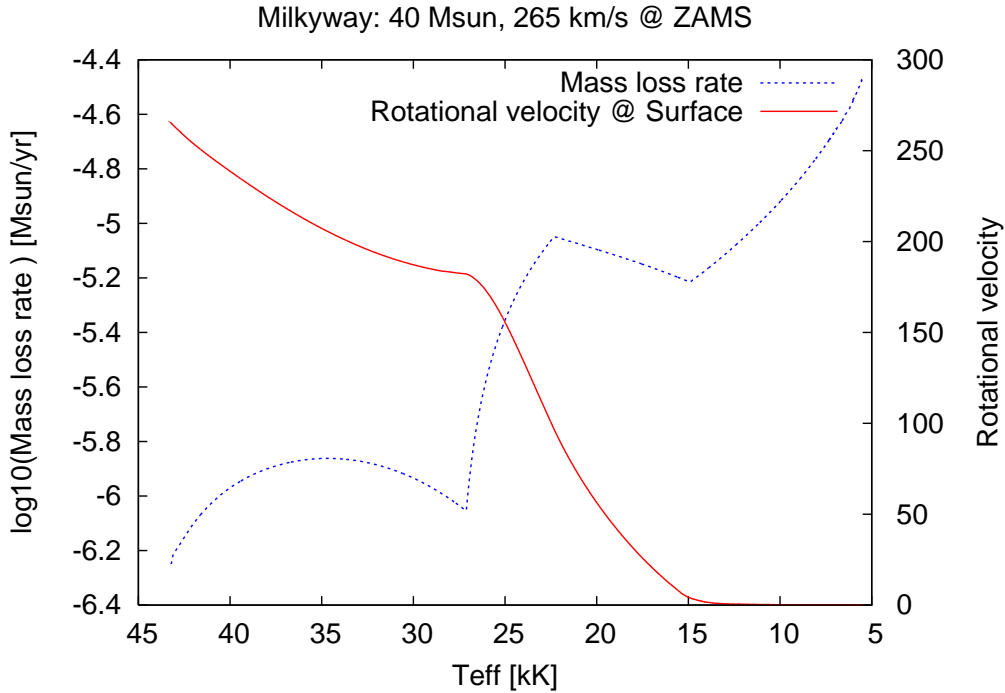}
  \caption{Mass-loss rate (dotted line) and rotational velocity (solid line) of a Galactic 40 \msun\ star 
which had a initial rotational velocity of 265 \kms\ on the ZAMS. 
A rather large overshootig parameter of 0.335 of a pressure scale-height was employed. The mixing efficiency
of fc=0.0228 represents an efficiency factor with which the diffusion coefficients of the different 
rotational mixing processes are multiplied (see Hunter et al. 2008, Brott et al., in prep. for further details.)
Note that the second mass-loss increase below 15\,000 K is due to a swith from the Vink et al. (2000) recipe 
to a calibration by Nieuwenhuizen \& de Jager (1990) -- consistent with the position of the second bistability jump
predicted by Vink et al. and observed by Lamers et al. (1995) around spectral type A0.}
\label{fig:brott}
\end{figure}

Figure~\ref{fig:mp} shows a recent figure from Markova \& Puls (2008) 
showing how the rotational velocity of Galactic OB supergiants 
depends on spectral type. It can be noted that $v$sin$i$ drops from $\sim$100 \kms\ to $\sim$50 \kms\ 
close to spectral type B1 -- the position of the bi-stability jump (Lamers et al. 1995, 
Crowther et al. 2006).
As we are interested in checking whether the predicted jump in mass loss by a factor of
five at the bi-stability jump (Vink et al. 1999, 2000) might potentially explain the steep drop in rotation due 
to the loss of angular momentum evolutionary tracks were
computed with this in mind (Brott et al. in prep.). Figure~\ref{fig:brott}
shows both the Vink et al. (2000) mass-loss rates (dotted line) and the predicted rotational
velocity (solid line) of a Galactic 40 \msun\ star which had a initial rotational velocity of 265 \kms\ on the
zero-age main-sequence (ZAMS). In order for the angular momentum removal to be maximal, a rather
large overshooting parameter of 0.335 of a pressure scale-height was employed, as this
provides a long time interval on the MS for the mass loss to be most efficient.

As an aside, we note that the drop in rotational velocities at a specific temperature is 
reminiscent of a situation encountered for Horizontal Branch stars (e.g. Behr et al. 2000). 
Vink (2007a) recently reviewed whether the steep jump at 10\,000 K could be 
due to the onset of radiation-driven winds as a result of the higher metal content for objects warmer than the jump temperature. 

\section{Could the changing winds of Luminous Blue Variables change the evolutionary paradigm?}
\label{sec:lbv}

Above we considered the effects of the bi-stability jump for main-sequence stars evolving 
from hot to cool temperatures. However, even more dramatic effects might occur for objects 
that have already lost a large fraction of their initial mass, finding themselves in close proximity 
to the Eddington limit. These Luminous Blue Variables change their effective temperatures on 
a variety of timescales (Humphreys \& Davidson 1994, Vink 2008).
The micro variations are not noticably different from small-amplitude variations in 
other BA supergiants.
At the other extreme, we find the super-outburst of objects such as Eta Car, sometimes 
referred to as SN-impostors, when observed in other galaxies (Van Dyk et al. 2000, Maund et al. 2006). 
We note that only two of these super-outbursts have been identified in the Milky Way (the eruptions of 
Eta Car in the 19th and P Cyg in the 17th century). 
The mass-loss rates involved in these giant eruptions are of order 0.1 $\msunyr$ and are too large 
to be explained by line acceleration. However, continuum-driven winds may well be able 
to provide the necessary driving (Smith \& Owocki 2006).
Most typifying for the class are the S~Dor variations, where objects 
vary on timescales of years to decades.
When LBVs such as AG Car -- one of the prototypes -- change their 
radii on their S~Dor timescales, they show large mass-loss variations (Stahl et al. 2001). 
Such variable wind behaviour has qualitatively been reproduced by radiation-driven wind 
models of Vink \& de Koter (2002). 
We anticipate that this type of wind behaviour may result in a circumstellar medium consisting of concentric shells with varying 
densities, which may have ramifications for the end-points of massive stars. 
Kotak \& Vink (2006) suggested that the quasi-periodic modulations seen in the radio lightcurves of some supernovae (SNe), 
such as 2001ig (Ryder et al. 2004) and 2003bg (Soderberg et al. 2006) may indicate that LBVs could be the {\it direct}
progenitors of some SNe.

At first this seems to contradict stellar evolution calculations, which do not predict
LBVs to explode, and such a scenario was until recently considered ``wildly speculative'' (Smith \& Owocki 2006). 
However, the intruiging supernova 2006jc showed a giant eruption just 2 years prior to explosion 
(Pastorello et al. 2007, Foley et al. 2007) which may add some confidence to the Kotak \& Vink suggestion that LBVs might explode. 
There have been a number of other studies suggestting that LBVs may explode. Gal Yam et al. (2007) reported the detection of 
a most luminous progenitor of SN 2005gl. Although the properties 
of the potential progenitor star are consistent with that of an LBV, a hypergiant cannot be classified as an LBV 
until it has shown S~Dor or Eta-Car-type variability (Humphreys \& Davidson 1994, Vink 2008). 
Another interesting hint that LBVs may explode arises from the similarities 
in LBV nebula morphologies and the circumstellar medium of SN 1987A (Smith 2007). Finally, one the most 
luminous supernova ever recorded, SN 2006gy (e.g. Ofek et al. 2007) may also have been an Eta Carinae type LBV (Smith et al. 2007).

As current state-of-the-art stellar evolution calculations do not predict LBVs to explode, this 
represents a major unresolved problem in the physics of massive stars.
In the generally upheld picture for the evolution of the most massive stars, 
LBVs are considered ``transitional'' objects in a phase before 
entering the He-burning WR stage, by the end of which
the WR star is expected to explode as a type Ib/c supernova.
The reason for the common (Conti 1976) scenario:

\[ O \rightarrow {\rm LBV} \rightarrow
{\rm WR} \rightarrow{\rm SN}, \]
is that LBVs are He (and N) rich compared to O stars, yet H-rich (thus He-poor)  
compared to WR stars. The situation is even more complex
as there is also a group of high-luminosity late-type WR stars which are H-rich and seem 
closely related to the classical LBVs in quiescence. 
In particular, we note that the R127 was a late-type WN star (Of/WN9), before it went into outburst.
Nonetheless, the picture of a relatively short-lived, some $10^{4}$ yrs, 
core H-burning LBV phase prior to a more extensive spell of a few times $10^{5}$ yrs
core He-burning WR phase seemed well established -- until recently.

\begin{figure}
 \includegraphics[height=.6\textheight,angle=90]{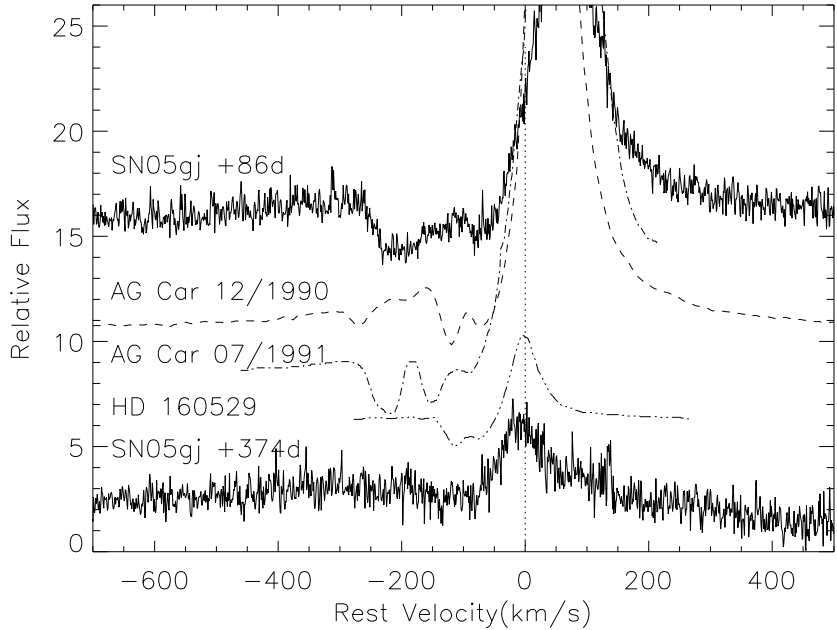}
\caption{Multiple absorption components seen in the P Cygni H$\alpha$ profile 
of SN 2005gj (top) in comparison to the LBVs AG~Car and HD\,160529.
The figure has been taken from Trundle et al. (2008).} 
\label{fig:trundle}
\end{figure}

Whilst the quasi-sinusoidal modulations in the radio lightcurves of transitional SNe may 
possibly also be explained by alternative scenarios\footnote{Although there are other explanations for 
these radio modulations, none of these are entirely satisfactory. Ryder et al. (2004) suggested the modulations might be 
due to a WR pinwheel system where a secondary star perturbs the circumstellar medium of the primary WR star. 
Although this remains possible (though 
the inferred radial spacings in Ryder et al. (2004) are incorrect by a factor of 10, see Kotak \& Vink 2006), the 
fact that SN 2003bg is so similar to 
SN 2001ig led Soderberg et al. (2006) to suggest the modulations are more likely due to a variable WR wind of a single 
star, resulting in concentric shells. However, such variability has never been observed in WR stars. This shortcoming 
was alleviated with the S\,Dor LBV suggestion of Kotak \& Vink. 
Finally, we mention the possibility that the radio modulations 
might be due to a variable wind of a pulsating red supergiant (RSG; Heger et al. 1997), however 
the problem with such a scenario is that a RSG is H-rich, whilst SN 2003bg was first classified as a SN Ic. The fact that 
SNe 2001ig and 2003bg showed exactly the opposite transitional behaviour between type I and II, or H-rich or H-poor, was 
an extra reason for Kotak \& Vink to consider LBVs as possible progenitors, as LBVs are H-rich compared to WR stars, but
H-poor compared to RSGs.}, it might be relevant that the 
{\it same} underlying mechanism , i.e. wind bi-stability, might 
account for wind-velocity variations seen spectroscopically in the SN 
SN 2005gj (Trundle et al. 2008).
Here, the variable winds are inferred from double P~Cygni components (see Fig.~\ref{fig:trundle}) 
which appear almost identical to those seen in the H$\alpha$ profiles of the well-known 
S\,Dor variables AG\,Car and HD\,160529. It should also be noted that the timescales and the 
spectroscopically measured wind velocities of SN 2005gj, with $\vinf$ $\simeq$100-200 \kms, 
are consistent with those of LBVs, whilst they are yet again not consistent with those 
of the much slower RSG winds ($\sim$10 \kms), or the much faster WR winds ($\simeq$1000-5000 \kms).

\section{Mass loss as a function of metallicity}
\label{sec:z}

\begin{figure}
 \includegraphics[height=.4\textheight]{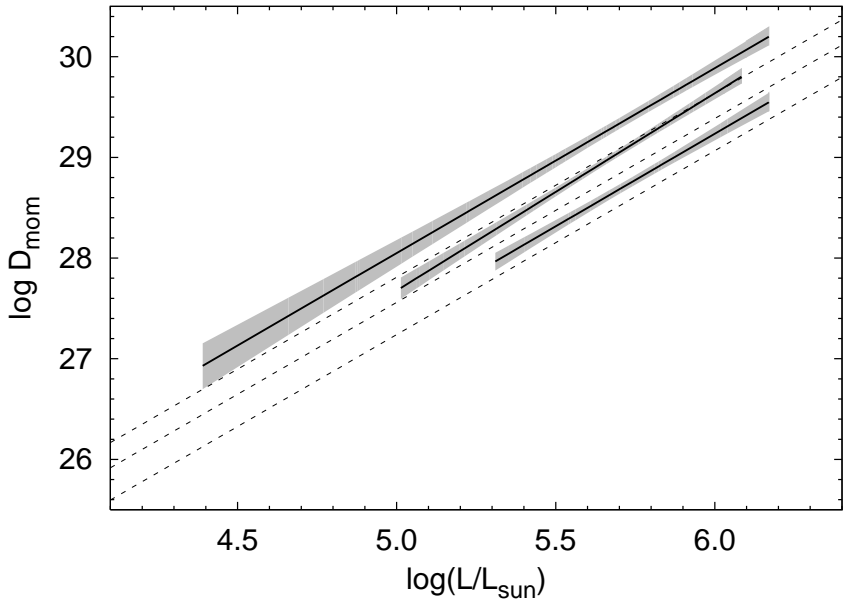}
\caption{This figure represents a confrontation between O star mass-loss predictions (dashed line) and 
recent empirical mass-loss rates in the form of the so-called wind momentum-luminosity relationship for Galactic (top), 
LMC (middle) and SMC (bottom) O stars. The figure is from Mokiem et al. (2007).}
  \label{f:wlr_comp}
\end{figure}

Metallicity is a key parameter in the physics of stars and star-forming galaxies, largely 
via metallicity-dependent stellar winds. We compare the predictions of O star 
mass-loss rates (Vink et al. 2001) with recent empirical mass-loss rates from 
Mokiem et al. (2007) using the so-called wind momentum-luminosity relationship (Kudritzki \& Puls 2000) 
for Galactic, Large Magallanic Cloud, and Small Magallanic Cloud O stars in the $Z$ range from solar 
to only 20\% solar. We note that for all three galaxies, the empirical rates are 
somewhat larger (by a factor $\sim$2) than predicted by Vink et al. (see Fig.\ref{f:wlr_comp}). 
As the empirical rates are most likely affected by wind clumping (e.g. Bouret et al. 2003, Martins et al. 2005, 
Hamann et al. 2008), the 
empirical rates are likely maximal. Therefore, when we assume a modest clumping factor corresponding to an 
empirical $\dot{M}$ reduction of a factor $\sim$2, the empirical rates show very good agreement with theory. 
The wind clumping factor however remains an unsolved problem in stellar astrophysics and if the true wind clumping is larger than assumed, with 
empirical $\dot{M}$ overestimates of $\sim$10 as some studies suggest (e.g. Fullerton et al. 2006), 
the current mass-loss predictions might also be too large. This is certainly an important topic for future investigation.

\begin{figure}
 \includegraphics[height=.4\textheight]{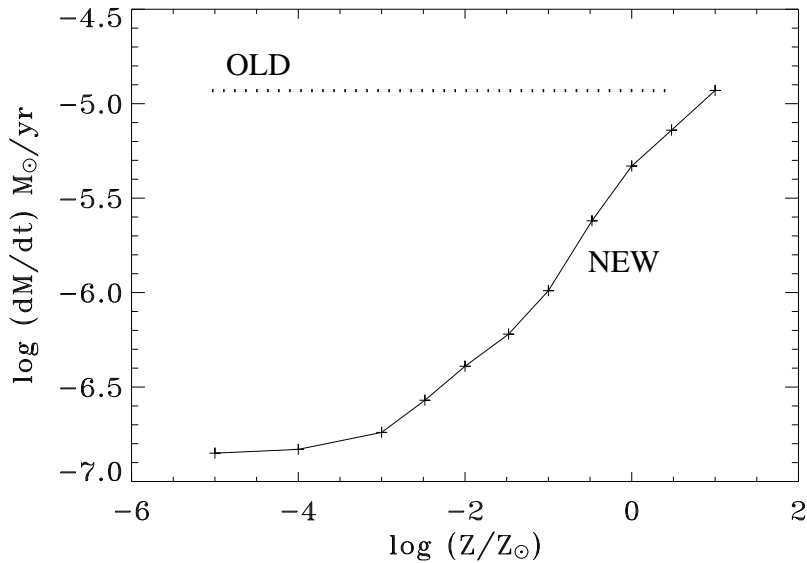}
 \caption{$Z$-dependent mass-loss predictions for Wolf-Rayet stars (Vink \& de Koter 2005).
Albeit the overwhelming presence of carbon for all $Z$, the new WR mass-loss rate does {\it not} show
$Z$-independent behaviour as assumed previously (dotted line).
The new computations show that WR mass loss depends strongly on iron ($Z$) --- a {\it key}
result for predicting a high occurrence of long-duration GRBs at low $Z$.}
  \label{f:mdotz2}
\end{figure}

Massive stars lose mass at even higher rates during the more evolved LBV and WR phases. 
During the latter phase, the outer layers become strongly chemically enriched, which may potentially 
modify mass loss through winds. 
Vink \& de Koter (2005) investigated the mass loss versus $Z$ dependence for late-type
WR stars using a Monte Carlo approach (see Fig.\ref{f:mdotz2}). 
Despite the overwhelming presence of carbon at all $Z$, $\dot{M}$ does {\it not} show 
a $Z$-independent behaviour (as was generally assumed previously), but WR mass loss depends strongly on the iron (Fe) opacity, just 
like for O stars (Vink et al. 1999, Gr\"afener \& Hamann 2008).

Furthermore, although the \mdot\ versus $Z$ dependence is consistent with a power-law decline in the observable 
Universe down to log $Z/\zsun \sim -3$, it flattens off for extremely low $Z$ models.
The reason is that carbon, nitrogen, oxygen, hydrogen and helium 
take over the driving from Fe which dominates the higher $Z$ domain.
The strong $Z$-dependence of WR winds where the WR $\dot{M}$ drops by orders of magnitude, 
might represent a key result for the high incidence of long-duration GRBs at low 
metallicity. 
The favoured progenitors of long GRBs are thought to be rapidly rotating 
WR stars. However, most Galactic WR stars are slow rotators, as stellar winds probably 
remove the necessary stellar angular momentum, potentially posing a challenge to the collapsar model for 
GRBs.

Observational data however indicate that GRBs occur predominately in low metallicity ($Z$) galaxies (e.g. Le Floc'h et al. 2003, Prochaska et al. 2004, Vreeswijk et al. 2004, Modjaz et al. 2008), which may 
resolve the problem: lower $Z$ leads to less mass loss, which may inhibit  
angular momentum removal, allowing WR stars to remain rotating rapidly until 
collapse (Yoon \& Langer 2005, Woosley \& Heger 2006).
As a test of this scenario, 
Vink (2007b) performed a linear spectropolarimetry survey of WR stars in the low $Z$ 
environment of the LMC and found an incidence of line polarisation effects in LMC WR stars as low as that of the Galactic sample of 
Harries et al. (1998). This suggests that the threshold metallicity where significant 
differences in WR rotational properties occur is below that of the LMC (at $Z$ $\sim$ 0.4 $\zsun$), possibly constraining GRB 
progenitor channels to this upper metallicity.

\section{Conclusions}
\label{sec:concl}

We presented theoretical mass-loss rates and their implications for the peeling off and angular momentum loss 
of massive stars during the various evolutionary stages from the main sequence, to the LBV, and WR phases.
The role of the bi-stability jump at an effective temperature of $\sim$25\,000 K was discussed in the context 
of the observed drop in rotational velocities in this part of the Hertzsprung-Russell diagram as well as 
for the variable winds of LBVs, suggesting these objects could be in a {\it direct} pre-SN state of massive 
star evolution.

\begin{acknowledgments}
\noindent
We would like to thank all our collegues and friends in the field of massive stars and in particular to 
Alex de Koter and Rubina Kotak. Special thanks also to Ines Brott and Norbert Langer for computing the 
evolution of the rotation rate with time and temperature.
\end{acknowledgments}

\end{document}